# Precursor Selection in Hybrid Molecular Beam Epitaxy of Alkaline-Earth Stannates

Abhinav Prakash[1,†], Tianqi Wang[1,†], Rashmi Choudhary[1], Greg Haugstad[2], Wayne L. Gladfelter[3], and Bharat Jalan[1,*]

[1]Department of Chemical Engineering and Materials Science, University of Minnesota, Minneapolis, Minnesota 55455, USA

[2]Characterization Facility, University of Minnesota, Minneapolis, Minnesota 55455, USA

[3]Department of Chemistry, University of Minnesota, Minneapolis, Minnesota 55455, USA

[†]Both authors contributed equally to this work
[*]Corresponding author: bjalan@umn.edu

# Abstract

One of the challenges of oxide molecular beam epitaxy (MBE) is the synthesis of oxides containing metals with high electronegativity (metals that are hard to oxidize). The use of reactive organometallic precursors can potentially address this issue. To investigate the formation of radicals in MBE, we explored three carefully chosen metal-organic precursors of tin for $SnO_2$ and $BaSnO_3$ growth: tetramethyltin (TMT), tetraethyltin (TET), and hexamethylditin (HMDT). All three precursors produced single-crystalline, atomically smooth, and epitaxial $SnO_2$ (101) films on $r$-$Al_2O_3$ ($10\bar{1}2$) in the presence of an oxygen plasma. The study of growth kinetics revealed reaction-limited and flux-limited regimes except for TET, which also exhibited a decrease in deposition rate with increasing temperature above ~800°C. Contrary to these similarities, the performance of these precursors was dramatically different for $BaSnO_3$ growth. TMT and TET were ineffective in supplying adequate tin whereas HMDT yielded phase-pure, stoichiometric $BaSnO_3$ films. Significantly, HMDT resulted in phase-pure and stoichiometric $BaSnO_3$ films even without the use of an oxygen plasma (i.e., with molecular oxygen alone). These results are discussed using the ability of HMDT to form tin radicals and therefore, assisting with Sn → $Sn^{4+}$ oxidation reaction. Structural and electronic transport properties of films grown using HMDT with and without oxygen plasma are compared. This study provides guideline for the choice of precursors that will enable synthesis of metal oxides containing hard-to-oxidize metals using reactive radicals in MBE.

# Introduction

Molecular beam epitaxy (MBE) technique is a promising route to control defects and stoichiometry.[1-6] Most recently, oxide MBE has attracted attention for its ability to synthesize films of a metal oxide semiconductor (La-doped $BaSnO_3$, BSO) having the highest reported room temperature mobility (>180 $cm^2/Vs$[7]). These films were grown in adsorption-controlled mode using ozone-assisted MBE, which employs effusion cells for Ba and Sn, and an ozone source for oxygen. Similarly, a modified MBE approach utilizing radio-frequency (rf) oxygen plasmas was devised resulting in epitaxial growth of BSO films.[8] In both approaches, Sn was supplied in its pre-oxidized state in addition to employing reactive oxidants such as oxygen plasma or ozone to obtain complete oxidation of Sn → $Sn^{4+}$. While a pre-oxidized source ($SnO_2$ or mixture of $SnO + SnO_2$ for Sn) works, it may produce a low and unstable flux.[8] The use of elemental tin on the other hand resulted in either no film growth at higher temperatures or films with poor crystalline quality.[8] This aspect of BSO synthesis raises potentially a serious question on the efficacy of conventional/modified MBE for metal-oxide growth of stubborn (hard-to-oxidize) metals if a suitable pre-oxidized source material is not available. In principle, higher oxidant pressures can be used but this can lead to undesirable consequences in MBE, such as metal flux instability due to the source surface oxidation, or filament oxidation and even a low mean free path.[3-5, 9, 10]

Hybrid MBE approaches[11-14] (traditionally known as metal-organic or organometallic MBE[15, 16]) can potentially overcome the issue of incomplete oxidation to a certain extent, by supplying elements in the form of chemical precursors with metal atoms already bound to oxygen (in a pre-oxidized state albeit having high vapor pressure).[13] For example, titanium tetraisopropoxide (TTIP)[11-14] and vanadium oxytriisopropoxide (VTIP)[17] precursors are used to supply Ti and V. The advantages of this approach include scalable growth rates, self-regulating stoichiometry control, and potentially superior oxygen stoichiometry because of the additional source of oxygen from precursor.[4, 10] However, it was recently shown that use of an oxygen-containing metal precursor for tin (tin tetra-t-butoxide) to assist the growth in this fashion was unsuccessful.[18] An alternative to this approach is to supply tin in the form of volatile organometallic compounds. Monometallic tin alkyls, including TMT and TET have been shown to be effective tin precursors in chemical vapor and atomic layer depositions of $SnO_2$.[19-21] We recently demonstrated that the use of hexamethyditin (HMDT) as a tin source in MBE can yield phase-pure, epitaxial BSO using rf-oxygen plasma. Self-regulating cation stoichiometry using adsorption-controlled growth was identified resulting in films with exceptionally high conductivity exceeding $10^4$ Scm$^{-1}$ with high electron mobility at room temperature.[18, 22-24]

In this paper, we investigate the effectiveness of HMDT as a tin source for alkaline-earth stannate by performing detailed study of growth kinetics, and then comparing it with other precursors. We provide specific guidelines for selecting precursors that can form reactive radical intermediates. The general rule of thumb for precursor selection in hybrid oxide MBE includes precursors that are thermally stable, volatile, and free of halogen (F, Cl, etc. to avoid the formation of corrosive by-products). One should also check for by-product volatility, which is important to ensure they are pumped out effectively during growth. Furthermore, special attention should be paid to the choice of vacuum pumps and the nature of chemical precursor in hybrid MBE system. For example, if a turbo pump with a *high compression ratio* is used, it could be susceptible to potential damage as precursor vapor can condense into a liquid phase and damage the turbo pump.

Based on these criteria, we identified for evaluation three Sn-containing metal-organic precursors (see Table 1)[25-27]: (1) tetramethyltin (($CH_3$)$_4$Sn, TMT), (2) tetraethyltin (($C_2H_5$)$_4$Sn or TET), and (3) HMDT ($CH_3$)$_3$Sn-Sn($CH_3$)$_3$).

## **Experiment**

**Thin film growth**

Phase-pure $SnO_2$ and BSO films were grown on *r*-plane $Al_2O_3$ ($10\bar{1}2$) and $SrTiO_3$ (STO) (001) substrates, respectively using an oxide MBE system (EVO 50, Scienta Omicron, Germany) with a base pressure of less than $1\times10^{-10}$ Torr. An rf-oxygen plasma source (Mantis, UK) was used to supply oxygen at a pressure of $5\times10^{-6}$ Torr. The plasma source was equipped with charge deflection plates operated at 250 V to prevent charged ions from reaching the substrates. Barium was supplied through a conventional low-temperature effusion cell. Three different metal-organic precursors were used for Sn – TMT, TET, and HMDT (Sigma-Aldrich, USA). Substrates were cleaned using oxygen plasma operated at 250 W for 20 min before growth to remove any undesired adsorbed carbon on the surface. Sn precursors were introduced via a gas injector (E-Science Inc., USA) kept at 60 °C. Precursors were thermally evaporated from a stainless-steel bubbler (T = 60 °C) with gas lines kept at 60-70 °C to prevent any redeposition. Sn precursor flux was controlled by regulating a linear leak valve using feedback from a Baratron® capacitance manometer. For the growth of $SnO_2$ films, a manometer pressure of 140 mTorr was chosen for all precursors. Given the high volatility of these precursors, no carrier gas was used. $SnO_2$ films were grown at substrate temperatures between 300 °C and 900 °C (thermocouple temperature) using an oxygen plasma, whereas a substrate temperature of 800 °C and 900 °C were used for BSO growth. For all samples grown at substrate temperature > 500 °C, substrates were cooled down in oxygen plasma until the temperature reached below 500 °C. For the growth of BSO films on STO (100) substrates, Ba beam equivalent pressures (BEPs) were varied between $9\times10^{-9}$ and $5\times10^{-8}$ Torr.

**Structural Characterization of Thin Films**

Reflection High-Energy Electron Diffraction (RHEED) from Staib Instruments was used to monitor the growth *in situ*. *Ex situ* structural characterization was carried out using a high-resolution Philips Panalytical X'Pert thin film diffractometer using Cu $K_\alpha$ radiation. X-ray diffraction 2θ-ω coupled scans were performed to examine the phase purity and to measure the out-of-plane lattice parameter ($a_{op}$) of thin films. GenX software was used to fit the X-ray reflectivity to determine the films thicknesses.[28] Film surface morphology was studied using Atomic Force Microscopy (AFM) in contact mode. Rutherford backscattering spectrometry (RBS) was employed to quantify the cation stoichiometry (Sn to Ba ratio). A 4.278 MeV $He^{2+}$ beam was used to acquire the spectrum. RBS spectra were later fitted using Gaussian peaks. Details of the RBS measurements can be found in ref.[24]

**Electrical Characterization of La-doped BaSnO$_3$ Films**

Electronic transport measurements were carried out in a conventional van der Pauw geometry using indium as an Ohmic contact in Quantum Design Physical Properties Measurement System (PPMS® DynaCool™). Prior to transport measurements, films were annealed in a rapid thermal annealer (RTA) at 800 °C for 2 min. An oxygen flow rate of 10 sccm was used. No measurable conduction was observed in substrate.

## **Results and Discussion**

Figure 1 shows the equilibrium vapor pressure of TET, TMT, HMDT, and elemental Sn as a function of temperature showing orders of magnitude higher vapor pressure for precursors as compared to elemental solid Sn. Figures 2a-c show on-axis high-resolution 2θ-ω coupled scans for representative $SnO_2$ films on *r*-$Al_2O_3$ (10$\bar{1}$2) grown using TMT, HMDT, and TET. Substrate temperature and oxygen plasma pressure were kept at 700 °C and $5 \times 10^{-6}$ Torr, respectively. All precursors yielded epitaxial films consistent with single crystalline $SnO_2$ (101) revealing their competency as a tin source. Figures 2d-f further shows the grazing incidence x-ray reflectivity scans for these films with clear Kiessig fringes suggesting uniform film thicknesses and smooth surface morphology. To this end, it may be persuasive to consider these precursors effective in supplying Sn for BSO growth. To examine the efficacy of these precursors as a tin source for BSO growth, we performed systematic growth kinetics study of $SnO_2$ film on $r$-$Al_2O_3$ ($10\bar{1}2$) substrates using TMT, HMDT, and TET. Sn flux was fixed by keeping Baratron® pressure at 140 mTorr for each precursor. Oxygen plasma pressure was kept at $5 \times 10^{-6}$ Torr.

Figure 3 shows growth rate as a function of substrate temperature between 300°C and 900°C. At low temperatures (< ~ 400 °C for TET, < ~ 500 °C for HMDT and < ~ 600 °C for TMT), the growth kinetics follow an Arrhenius behavior (reaction-limited regime). At intermediate temperatures (400 °C - 800 °C for TET, 500 °C - 900 °C for HMDT and 600 °C - 900 °C for TMT), growth rate remained temperature-independent (flux-limited regime). At T > 800 °C, growth rate decreased for TET (desorption-limited regime) similar to what is reported previously.[29] No decrease in growth rate was observed for films grown using HMDT or TMT. The ethyl ligands of TET apparently open a competing reaction path leading to desorption of a volatile Sn- or SnO-containing species, which results in a decreased growth rate at high temperatures.[29-33] It is further important to note that despite using the same Baratron® pressure setpoint in the depositions, TET resulted in higher growth rates than either HMDT or TMT at temperatures below 800 °C. Among the three precursors, TMT exhibited the lowest deposition rates. These observations indicate an important, quantitative impact of precursor composition on the deposition process. While future study should be directed to investigate the origin of these differences, it is clear that TMT and HMDT are more suitable candidates for tin for the growth of ternary oxides in an oxide MBE where higher growth temperature is preferred for better crystalline quality.

After studying the growth kinetics for $SnO_2$ deposition, we examined their effectiveness for BSO growth. Figure 4a shows the comparison of the RBS results of BSO film grown using HMDT and TMT as a Sn source. Ba BEP ($5 \times 10^{-8}$ Torr) and oxygen pressure ($5 \times 10^{-6}$ Torr) were kept identical between two growths whereas the Sn Baratron® setpoint of 200 mTorr was used for films grown using TMT and a setpoint of 75 mTorr was used for films grown using HMDT. BSO films using TMT resulted in significantly lower Sn incorporation (Sn : Ba = 0.14) whereas HMDT showed a nearly 1:1 Sn to Ba ratio. It is important to note that TMT seems ineffective as a source of tin even though a much higher BEP was used. Similarly, a comparison between TET and HMDT resulted in films with significantly lower Sn incorporation for TET. This is illustrated in figure 4b which shows RBS spectra for BSO films on STO (001) grown using HMDT and TET. Ba BEP, oxygen pressure and Sn Baratron® setpoints were fixed at $1\times10^{-8}$ Torr, $5\times10^{-6}$ Torr, and 200 mTorr, respectively. Significantly, the use of HMDT resulted in much higher Sn : Ba ratio = $2.82 \pm 0.02$, whereas TET yielded Sn-deficient films with Sn:Ba ratio = $0.63 \pm 0.02$. The source of these substantial precursor effects on the BSO film composition and quality remains unknown. It is possible that TMT and TET undergo competing reactions that lead to volatile Sn- or SnO-containing intermediates causing to the observed Sn deficiency. Based on experimental and theoretical studies, the Sn-Sn bond enthalpy in HMDT is ~ 35 kJ/mol smaller than the Sn-C bond in TMT.[34-36] More facile cleavage of the Sn-Sn bond may lead to higher concentrations of surface-bound tin radicals and thus overcome the impact of competing desorption reactions.

To further investigate the effectiveness of HMDT towards oxidation reaction (Sn $\rightarrow$ $Sn^{4+}$) under standard MBE environment, we performed BSO growth with molecular oxygen. Significantly, HMDT yielded phase-pure, stoichiometric BSO films with molecular oxygen providing a strong evidence of reactive radical formation in agreement with the prior study.[34] Figure 5 shows on-axis high-resolution 2θ-ω coupled scans of BSO films grown using oxygen plasma and molecular oxygen. While both yielded epitaxial, phase-pure, and stoichiometric films, films grown using molecular oxygen were rougher and had higher structural disorder as evident from their large rocking curve full-width-half-maxima (FWHM) (0.06° for film grown with oxygen plasma vs. 1.90° for film grown using molecular oxygen). Interestingly, the bulk-like out-of-plane lattice parameter ($a_{op}$ = 4.116 ± 0.002 Å) was achieved for a 42 nm thick BSO films grown using molecular oxygen, whereas even a thicker film (92 nm) grown using oxygen plasma resulted in partially strained film on STO substrates with a slightly expanded $a_{op}$ = 4.127 ± 0.002 Å suggesting noticeably different relaxation mechanisms. It is conceivable that films grown using molecular oxygen possess higher amount of oxygen vacancies and related defects, which may assist in strain relaxation. Future investigations should be directed towards investigating strain relaxation behavior, microstructure and ionic conduction (for defect characterization) in these films.

Figures 6a and 6b show Hall mobility and electron density in La-doped BSO/STO (001) grown using oxygen plasma, and molecular oxygen, respectively. Stoichiometric composition determined using RBS is marked by blue- and red-shaded regions, respectively, in figures 6a and 6b. Mobility in excess of 100 $cm^2V^{-1}s^{-1}$ was found for oxygen plasma grown stoichiometric films whereas stoichiometric films grown using molecular oxygen (see figure S1) yielded much lower mobility (~ 16 $cm^2V^{-1}s^{-1}$) despite similar electron density. Furthermore, films grown using oxygen plasma showed stronger dependence of electron mobility and carrier density on non-stoichiometry. Much weaker dependence on stoichiometry was however found for films grown using molecular oxygen. While the exact reason(s) for the lower mobilities and its weak dependence on non-stoichiometry remains to be further investigated, these results clearly attest to the viability of radical-based MBE approach of metal oxide growth of stubborn metals using reactive radicals.

# Summary

In this paper, we have investigated three precursors TMT, TET, and HMDT as a Sn source for BSO films. All three precursors produced epitaxial, phase-pure $SnO_2$ films on $r$-$Al_2O_3$ with uniform and smooth surface morphology. Yet, growth kinetics study revealed an absence of desorption-limited regime for $SnO_2$ films grown using HMDT and TMT. Remarkably, HMDT resulted in epitaxial, phase-pure, stoichiometric BSO films using molecular oxygen, which has not been possible thus far due to stubbornness of Sn towards oxidation under the nominal MBE growth conditions. Hall mobility and carrier density of BSO films grown using molecular oxygen was found to be quite insensitive to non-stoichiometry-related defects in contrast to those grown using oxygen plasma. This study opens up new possibilities for MBE growth of metal oxides containing stubborn metals using reactive radicals in ultra-high vacuum techniques. It further informs future work towards the design and synthesis of functional precursors with radical chemistry.

# Acknowledgements

This work was supported primarily by the U.S. Department of Energy through DE-SC0020211. AP and TW acknowledge support from the University of Minnesota Doctoral Dissertation Fellowship. WG acknowledges support from the National Science Foundation under award number DMR-1607318. Portions of this work were conducted in the Minnesota Nano Center, which is supported by the National Science Foundation through the National Nano Coordinated Infrastructure Network (NNCI) under Award Number ECCS-2025124. Structural characterizations were carried out at the University of Minnesota Characterization Facility, which receives partial support from NSF through the MRSEC under Award Number DMR-2011401.

**Figures** (Color Online):

Table 1: Physical properties of tin precursors used for oxide growth using hybrid MBE

| Precursor | Tetramethyltin | Tetraethyltin | Hexamethylditin |
|---|---|---|---|
| Abbreviation | TMT | TET | HMDT |
| Chemical Formula | $(CH_3)_4Sn$ | $(CH_3CH_2)_4Sn$ | $(CH_3)_6Sn_2$ |
| Boiling Point | 75°C | 181°C | 182°C |
| Decomposition Temperature | 277°C | 180 °C | 250°C |

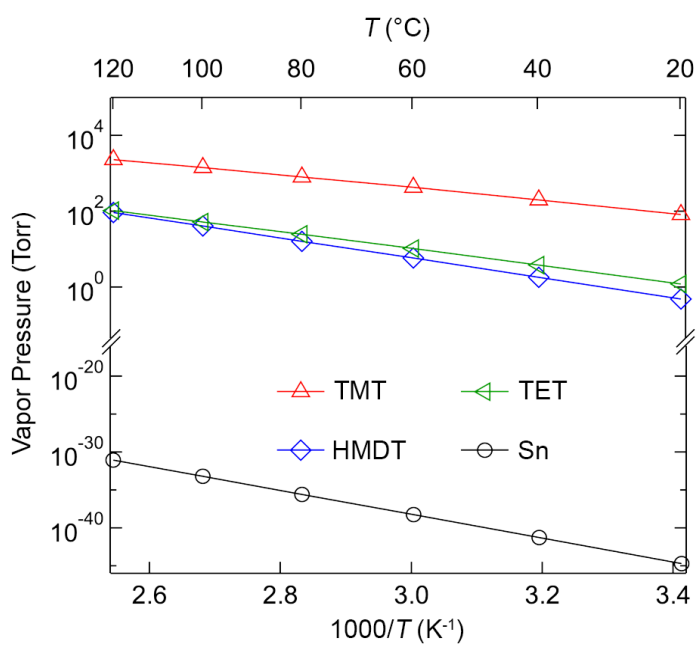

Figure 1: Equilibrium vapor pressures for the three precursors used in this study – tetramethyltin (TMT), hexamethylditin (HMDT), and tetraethyltin (TET). The vapor pressure curve for elemental Sn is also shown for comparison. Vapor pressure data taken from Refs.[37-39]

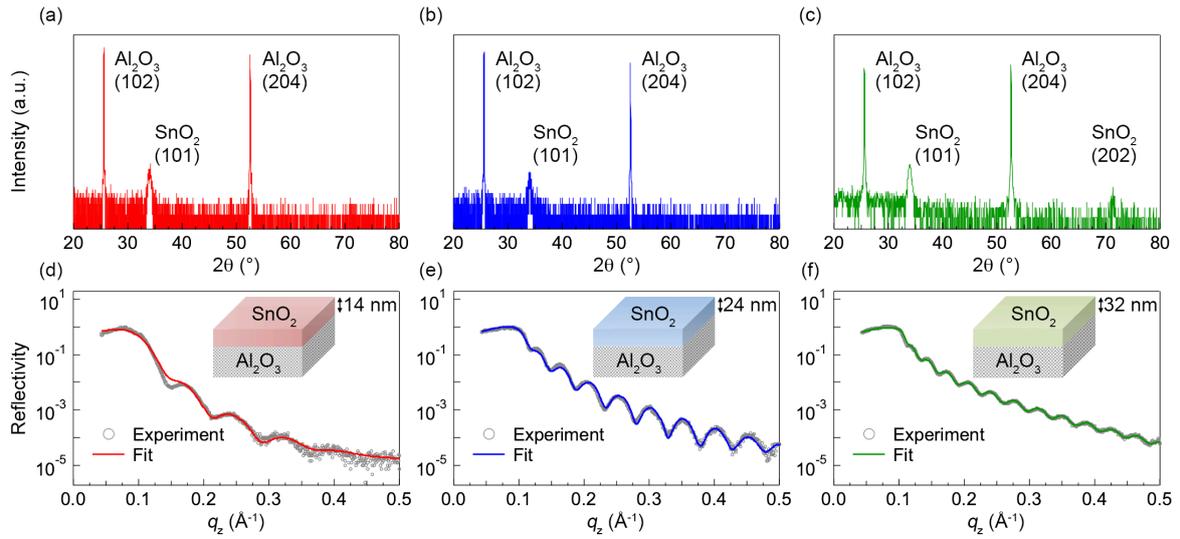

**Figure 2:** Wide-angle X-ray diffraction pattern (2θ-ω coupled scan) of SnO2 films grown on *r*-Al$_2$O$_3$ (10$\bar{1}$2) at a substrate (thermocouple) temperature of 700°C using **(a)** tetramethyltin (TMT), **(b)** hexamethylditin (HMDT), and **(c)** tetraethyltin (TET) in the presence of oxygen plasma operated at 250 W and a pressure of 5×10$^{-6}$ Torr. **(d-f)** Corresponding X-ray reflectivity used to determine the film thickness along with fits (solid lines) to the experimental data (symbols). Insets in each panel (d-f) show the schematics of the film structure and the extracted thickness values from fitting.

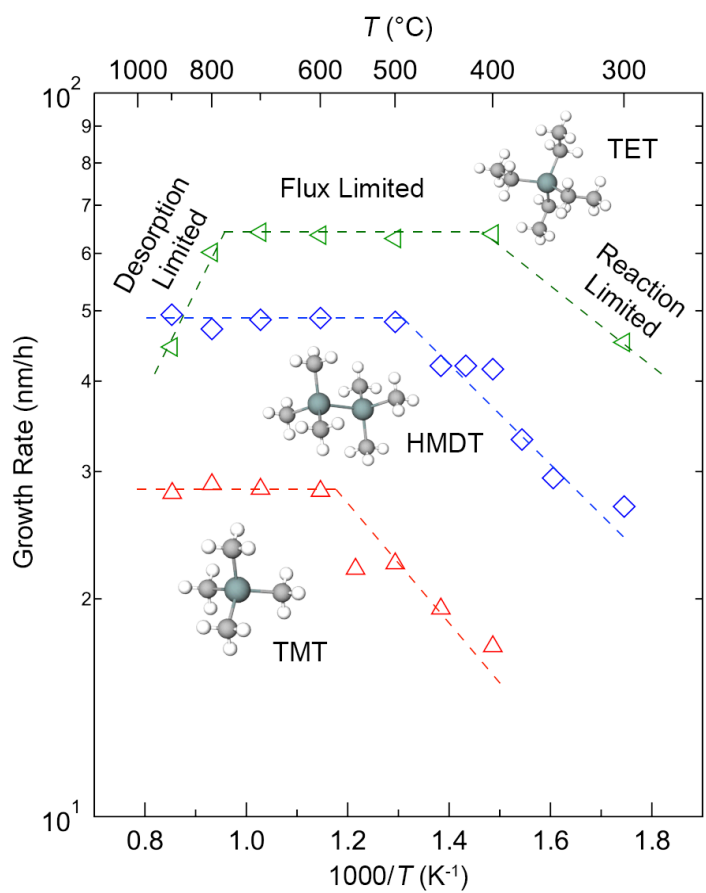

**Figure 3:** Growth rates of SnO$_2$ films on *r*-Al$_2$O$_3$ (10$\bar{1}$2) as a function of growth temperature for the three precursors – (red) tetramethyltin (TMT), (blue) hexamethylditin (HMDT), and (green) tetraethyltin (TET). All films were grown at a Baratron® setpoint pressure of 140 mTorr. Reaction-limited, flux-limited, and desorption-limited growth regimes are identified. The growth rates for SnO$_2$ films grown using TET are taken from ref.[29] Copyright 2015, American Vacuum Society.

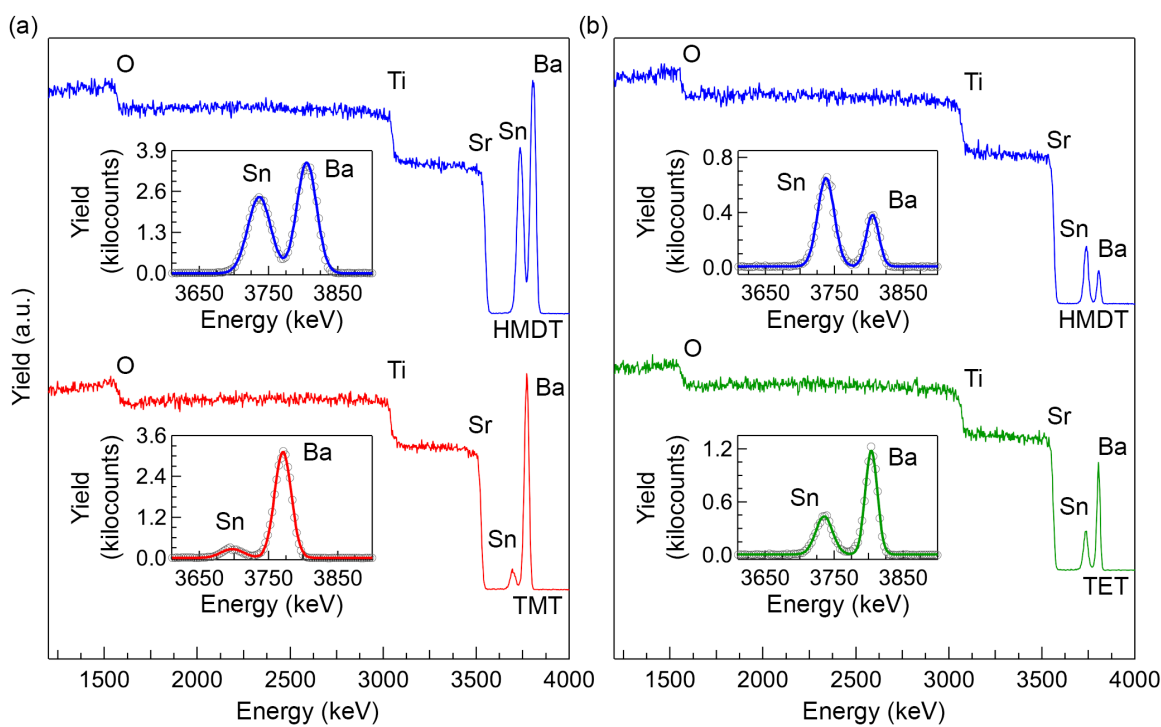

**Figure 4: (a)** Rutherford backscattering spectrometry (RBS) comparison for films grown using TMT (red) and HMDT (blue). Both films were grown at a substrate temperature of 900°C with a Ba BEP of $5\times10^{-8}$ Torr. Sn Baratron® setpoint of 200 mTorr was used for the film grown with TMT while a setpoint of 75 mTorr was used for the film grown with HMDT. Sn to Ba ratios calculated for the two films were $0.14 \pm 0.02$ (TMT) and $1.00 \pm 0.02$ (HMDT). **(b)** RBS comparison for the films grown with TET (green) and HMDT (blue). Both films were grown at a substrate temperature of 800 °C with a Ba BEP of $1\times10^{-8}$ Torr and Sn Baratron® setpoint of 200 mTorr. Sn to Ba ratios calculated for the two films were $0.63 \pm 0.02$ (TET) and $2.82 \pm 0.02$ (HMDT). Insets show the RBS data (symbols) around Sn and Ba spectra alongwith Gaussian fits (solid lines) used to determine the cation stoichiometry.

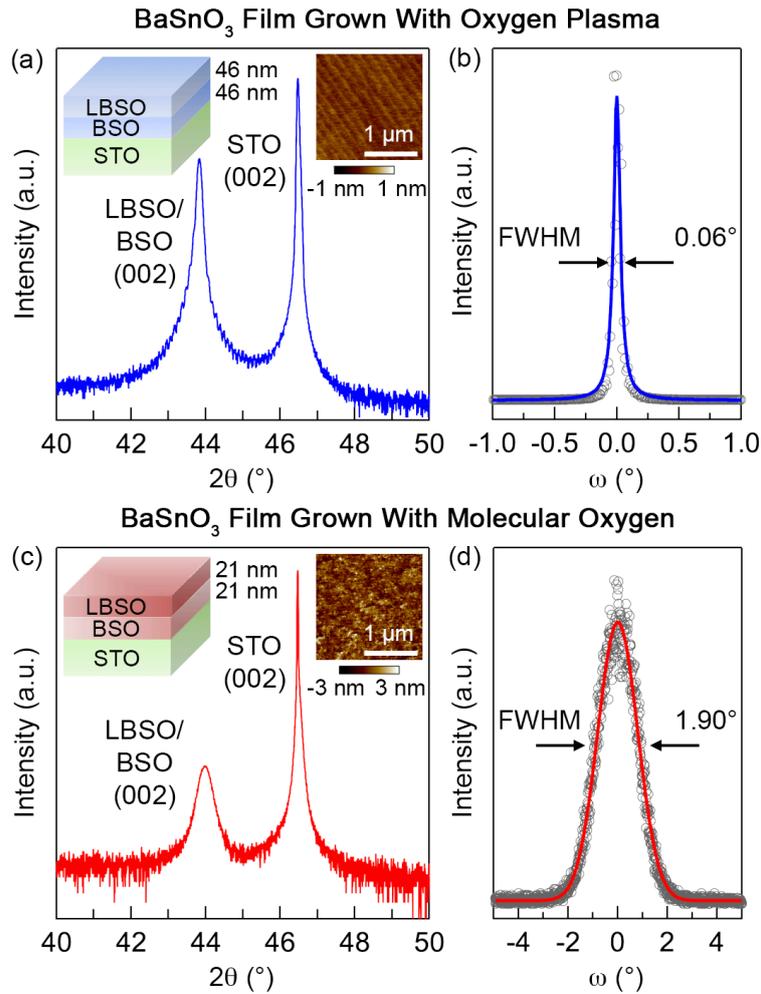

**Figure 5: (a)** Wide-angle X-ray diffraction and **(b)** rocking curve (ω scan) of a 46 nm LBSO/ 46 nm BSO/ STO (001) film grown using hexamethylditin (HMDT) (Baratron® setpoint of 90 mTorr) using oxygen plasma. Ba BEP of $5\times10^{-8}$ Torr was used. **(c)** Wide-angle X-raydiffraction and **(d)** rocking curves for a 21 nm LBSO/ 21 nm BSO/ STO (001) grown using molecular oxygen. Baratron® setpoint of 140 mTorr for HMDT, Ba BEP of $3\times10^{-8}$ Torr, and oxygen BEP of $5\times10^{-6}$ Torr were used for growth. The substrate temperature and La cell temperature were kept fixed at 900°C and 1180°C respectively for both films. Insets show the schematics of the film structures along with AFM of as-grown films. Root mean square roughness of 3.9 Å and 7.1 Å were measured for films grown with oxygen plasma and molecular oxygen, respectively.

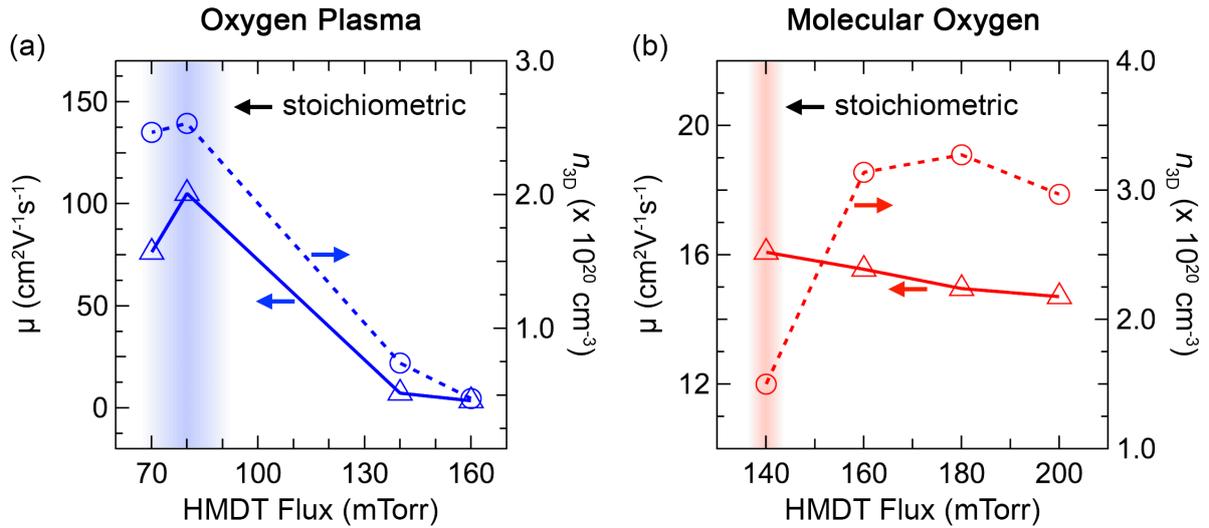

**Figure 6:** Electron mobility (left-axis) and carrier density (right-axis) as a function of HMDT flux of **(a)** 46 nm LBSO/ 46 nm BSO/ STO (001) films grown using oxygen plasma; **(b)** 21 nm LBSO/ 21 nm BSO/ STO (001) grown using molecular oxygen. Ba BEP of $5\times10^{-8}$ Torr, an oxygen pressure of $5\times10^{-6}$ Torr, and a La cell temperature of 1230 °C were used for films grown using oxygen plasma. Ba BEP of $3\times10^{-8}$ Torr, an oxygen pressure of $5\times10^{-6}$ Torr, and a La cell temperature of 1180 °C were used for films grown using molecular oxygen. All growths were carried out at a substrate temperature of 900 °C. Data for the films grown with oxygen plasma is taken from ref.[24] with permission from The Royal Society of Chemistry.